\documentclass[dvipsnames, letterpaper]{article} % DO NOT CHANGE THIS
\usepackage{aaai2026}  % DO NOT CHANGE THIS
\usepackage{times}  % DO NOT CHANGE THIS
\usepackage{helvet}  % DO NOT CHANGE THIS
\usepackage{courier}  % DO NOT CHANGE THIS
\usepackage[hyphens]{url}  % DO NOT CHANGE THIS
\usepackage{graphicx} % DO NOT CHANGE THIS
\usepackage{natbib}  % DO NOT CHANGE THIS AND DO NOT ADD ANY OPTIONS TO IT
\usepackage{caption} % DO NOT CHANGE THIS AND DO NOT ADD ANY OPTIONS TO IT
\usepackage{algorithm}
\usepackage{algorithmic}
\newcommand{\algorithmicprocedure}{\textbf{procedure}\ }
\newcommand{\PROCEDURE}[2]{\STATE \algorithmicprocedure \textsc{#1}(#2)}
\newcommand{\ENDPROCEDURE}{\STATE \textbf{end procedure}}
\usepackage[T1]{fontenc}
\usepackage{appendix}
\usepackage{mathtools,amssymb,amsfonts}
\usepackage{enumitem}
\usepackage{multicol}
\usepackage[dvipsnames]{xcolor}
\usepackage{tcolorbox}
\usepackage{booktabs}
\DeclareMathOperator*{\argmin}{arg\,min}

\usepackage{newfloat}
\usepackage{listings}
\usepackage{tikz,pgfplots}
\usepackage{pgfplots}
\usepackage{pgfplotstable}
\usepackage{tikz-3dplot}
\usetikzlibrary{shapes.geometric,backgrounds,patterns, trees}
\usetikzlibrary{3d,decorations.text,shapes.arrows,positioning,fit,backgrounds}
\usetikzlibrary{positioning, decorations.pathmorphing, shapes}
\usetikzlibrary{decorations.pathreplacing}
\usetikzlibrary{shapes.geometric,backgrounds,patterns, trees}
\usetikzlibrary{spy}
\usetikzlibrary{arrows.meta,
                bending,
                intersections,
                quotes,
                shapes.geometric}
            \usetikzlibrary{automata, positioning}
\definecolor{bluetwo}{RGB}{189, 213, 234}
\definecolor{bluethree}{RGB}{165, 193, 224}
\definecolor{gray2}{HTML}{ededed}
\definecolor{gray3}{HTML}{F5F5F5}
\definecolor{RoyalAzure}{rgb}{0.0, 0.22, 0.66}
\definecolor{lightgray}{gray}{0.9}
\definecolor{lightgray}{gray}{0.9}
\definecolor{lightgreen}{rgb}{0.88, 1, 0.88}
\definecolor{lightred}{rgb}{1, 0.88, 0.88}
\definecolor{lightblue}{rgb}{0.88, 0.94, 1}
\definecolor{lightorange}{rgb}{1, 0.94, 0.88}
\DeclareCaptionStyle{ruled}{labelfont=normalfont,labelsep=colon,strut=off} % DO NOT CHANGE THIS
\usepackage{colortbl}
\usetikzlibrary{shapes.geometric,backgrounds,patterns, trees}
\usetikzlibrary{3d,decorations.text,shapes.arrows,positioning,fit,backgrounds}
\usetikzlibrary{positioning, decorations.pathmorphing, shapes}
\usetikzlibrary{decorations.pathreplacing}
\usetikzlibrary{shapes.geometric,backgrounds,patterns, trees}
\usetikzlibrary{spy}
\usetikzlibrary{arrows.meta,
                bending,
                intersections,
                quotes,
                shapes.geometric}
              \usetikzlibrary{automata, positioning}
              \usepgfplotslibrary{fillbetween}
\usetikzlibrary{shapes,arrows}
\usetikzlibrary{arrows.meta}
\usetikzlibrary{positioning}
\tikzset{set/.style={draw,circle,inner sep=0pt,align=center}}
\usetikzlibrary{automata, positioning}
  \usetikzlibrary{shapes,shadows}
  \tikzstyle{abstractbox} = [draw=black, fill=white, rectangle,
  inner sep=10pt, style=rounded corners, drop shadow={fill=black,
  opacity=1}]
\tikzstyle{abstracttitle} =[fill=white]
\usetikzlibrary{calc,positioning,shapes.geometric}
\usetikzlibrary{arrows.meta,arrows}
\usetikzlibrary{matrix}
\tikzstyle{cblue}=[circle, draw, thin,fill=cyan!20, scale=0.8]
\tikzstyle{qgre}=[rectangle, draw, thin,fill=green!20, scale=0.8]
\tikzstyle{rpath}=[ultra thick, red, opacity=0.4]
\tikzstyle{legend_isps}=[rectangle, rounded corners, thin,
                       fill=gray!20, text=blue, draw]
\tikzstyle{legend_overlay}=[rectangle, rounded corners, thin,
                           top color= white,bottom color=green!25,
                           minimum width=2.5cm, minimum height=0.8cm,
                           pinegreen]
\tikzstyle{legend_phytop}=[rectangle, rounded corners, thin,
                          top color= white,bottom color=cyan!25,
                          minimum width=2.5cm, minimum height=0.8cm,
                          royalblue]
\tikzstyle{legend_general}=[rectangle, rounded corners, thin,
                          top color= white,bottom color=lavander!25,
                          minimum width=2.5cm, minimum height=0.8cm,
                          violet]
                        \usetikzlibrary{matrix}

\colorlet{myRed}{red!20}
\tikzset{
  rows/.style 2 args={/utils/temp/.style={row ##1/.append style={nodes={#2}}},
    /utils/temp/.list={#1}},
  columns/.style 2 args={/utils/temp/.style={column ##1/.append style={nodes={#2}}},
    /utils/temp/.list={#1}}}
\usetikzlibrary{backgrounds,calc,shadings,shapes.arrows,shapes.symbols,shadows}
\definecolor{switch}{HTML}{006996}
\makeatletter
\pgfkeys{/pgf/.cd,
  parallelepiped offset x/.initial=2mm,
  parallelepiped offset y/.initial=2mm
}
\pgfdeclareshape{parallelepiped}
{
  \inheritsavedanchors[from=rectangle] % this is nearly a rectangle
  \inheritanchorborder[from=rectangle]
  \inheritanchor[from=rectangle]{north}
  \inheritanchor[from=rectangle]{north west}
  \inheritanchor[from=rectangle]{north east}
  \inheritanchor[from=rectangle]{center}
  \inheritanchor[from=rectangle]{west}
  \inheritanchor[from=rectangle]{east}
  \inheritanchor[from=rectangle]{mid}
  \inheritanchor[from=rectangle]{mid west}
  \inheritanchor[from=rectangle]{mid east}
  \inheritanchor[from=rectangle]{base}
  \inheritanchor[from=rectangle]{base west}
  \inheritanchor[from=rectangle]{base east}
  \inheritanchor[from=rectangle]{south}
  \inheritanchor[from=rectangle]{south west}
  \inheritanchor[from=rectangle]{south east}
  \backgroundpath{
    \southwest \pgf@xa=\pgf@x \pgf@ya=\pgf@y
    \northeast \pgf@xb=\pgf@x \pgf@yb=\pgf@y
    \pgfmathsetlength\pgfutil@tempdima{\pgfkeysvalueof{/pgf/parallelepiped
      offset x}}
    \pgfmathsetlength\pgfutil@tempdimb{\pgfkeysvalueof{/pgf/parallelepiped
      offset y}}
    \def\ppd@offset{\pgfpoint{\pgfutil@tempdima}{\pgfutil@tempdimb}}
    \pgfpathmoveto{\pgfqpoint{\pgf@xa}{\pgf@ya}}
    \pgfpathlineto{\pgfqpoint{\pgf@xb}{\pgf@ya}}
    \pgfpathlineto{\pgfqpoint{\pgf@xb}{\pgf@yb}}
    \pgfpathlineto{\pgfqpoint{\pgf@xa}{\pgf@yb}}
    \pgfpathclose
    \pgfpathmoveto{\pgfqpoint{\pgf@xb}{\pgf@ya}}
    \pgfpathlineto{\pgfpointadd{\pgfpoint{\pgf@xb}{\pgf@ya}}{\ppd@offset}}
    \pgfpathlineto{\pgfpointadd{\pgfpoint{\pgf@xb}{\pgf@yb}}{\ppd@offset}}
    \pgfpathlineto{\pgfpointadd{\pgfpoint{\pgf@xa}{\pgf@yb}}{\ppd@offset}}
    \pgfpathlineto{\pgfqpoint{\pgf@xa}{\pgf@yb}}
    \pgfpathmoveto{\pgfqpoint{\pgf@xb}{\pgf@yb}}
    \pgfpathlineto{\pgfpointadd{\pgfpoint{\pgf@xb}{\pgf@yb}}{\ppd@offset}}
  }
}

\makeatletter
\tikzset{anchor/.append code=\let\tikz@auto@anchor\relax,
  add font/.code=%
    \expandafter\def\expandafter\tikz@textfont\expandafter{\tikz@textfont#1},
  left delimiter/.style 2 args={append after command={\tikz@delimiter{south east}
    {south west}{every delimiter,every left delimiter,#2}{south}{north}{#1}{.}{\pgf@y}}}}
\tikzstyle{sms} = [rectangle callout, draw,very thick, rounded corners, minimum height=20pt]
\makeatletter
\tikzset{anchor/.append code=\let\tikz@auto@anchor\relax,
  add font/.code=%
    \expandafter\def\expandafter\tikz@textfont\expandafter{\tikz@textfont#1},
  left delimiter/.style 2 args={append after command={\tikz@delimiter{south east}
    {south west}{every delimiter,every left delimiter,#2}{south}{north}{#1}{.}{\pgf@y}}}}
\tikzstyle{sms} = [rectangle callout, draw,very thick, rounded corners, minimum height=20pt]
\usetikzlibrary{positioning,calc}
\tikzstyle{block} = [rectangle, draw,
text width=10.5em, text centered, rounded corners, minimum height=4em]
\tikzstyle{line} = [draw, -latex]
\tikzset{
  mybackground51/.style={execute at end picture={
      \begin{scope}[on background layer]
        \draw[black, rounded corners=2ex, fill=gray2] (current bounding box.south west)
        rectangle (current bounding box.north east);
        \node[draw,fill=white,ellipse,anchor=west,inner sep=1pt,minimum width=1ex] at (current bounding box.north
        west){#1};
      \end{scope}
    }},
}
\tikzset{
  mybackground9/.style={execute at end picture={
        \begin{scope}[on background layer]
          \draw[black,fill=black!5,rounded corners=6ex] (current bounding box.south west)
                    rectangle (current bounding box.north east);
          \node[draw,fill=white,ellipse,anchor=west,inner sep=1pt,minimum width=4ex] at (current bounding box.north
                   west){#1};
        \end{scope}
    }},
}
\tikzset{
  mybackground13/.style={execute at end picture={
        \begin{scope}[on background layer]
          \draw[black, fill=gray2, rounded corners=4ex] (current bounding box.south west)
                    rectangle (current bounding box.north east);
          \node[draw,fill=white,ellipse,anchor=west,inner sep=1pt,minimum width=4ex] at (current bounding box.north
                   west){#1};
        \end{scope}
    }},
}
\tikzset{
  mybackground14/.style={execute at end picture={
        \begin{scope}[on background layer]
          \draw[black, rounded corners=2ex] (current bounding box.south west)
                    rectangle (current bounding box.north east);
          \node[draw,fill=white,ellipse,anchor=west,inner sep=1pt,minimum width=4ex] at (current bounding box.north
                   west){#1};
        \end{scope}
    }},
}

\tikzset{
  mybackground6/.style={execute at end picture={
        \begin{scope}[on background layer]
          \draw[black,rounded corners=1ex, line width=0.15mm] (current bounding box.south west)
                    rectangle (current bounding box.north east);
          \node[draw,fill=white,ellipse,anchor=west,inner sep=1pt,minimum width=4ex] at (current bounding box.north
                   west){#1};
        \end{scope}
    }},
}

\tikzset{
  mybackground11/.style={execute at end picture={
        \begin{scope}[on background layer]
          \draw[black, fill=Black!80!Sepia!9, rounded corners=6ex] (current bounding box.south west)
                    rectangle (current bounding box.north east);
          \node[draw,fill=white,ellipse,anchor=west,inner sep=1pt,minimum width=4ex] at (current bounding box.north
                   west){#1};
        \end{scope}
    }},
}

\tikzset{
  mybackground15/.style={execute at end picture={
        \begin{scope}[on background layer]
          \draw[black, fill=Black!80!Sepia!9, rounded corners=3ex] (current bounding box.south west)
                    rectangle (current bounding box.north east);
          \node[draw,fill=white,ellipse,anchor=west,inner sep=1pt,minimum width=4ex] at (current bounding box.north
                   west){#1};
        \end{scope}
    }},
}

\tikzset{
  mybackground12/.style={execute at end picture={
        \begin{scope}[on background layer]
          \draw[black, fill=Black!40!Emerald!30, rounded corners=3ex, line width=0.3mm] (current bounding box.south west)
                    rectangle (current bounding box.north east);
        \end{scope}
    }},
}
\tikzset{
  mybackground18/.style={execute at end picture={
      \begin{scope}[on background layer]
        \draw[black, fill=gray3, rounded corners=3.5ex] (current bounding box.south west)
        rectangle (current bounding box.north east);
        \node[draw,fill=white,ellipse,anchor=west,inner sep=1pt,minimum width=4ex] at (current bounding box.north
        west){#1};
      \end{scope}
    }}
}
\tikzset{
  mybackground180/.style={execute at end picture={
      \begin{scope}[on background layer]
        \draw[black, rounded corners=3.5ex] (current bounding box.south west)
        rectangle (current bounding box.north east);
        \node[draw,fill=white,ellipse,anchor=west,inner sep=1pt,minimum width=4ex] at (current bounding box.north
        west){#1};
      \end{scope}
    }}
}
\tikzset{
  mybackground19/.style={execute at end picture={
      \begin{scope}[on background layer]
        \draw[black, rounded corners=3.5ex] (current bounding box.south west)
        rectangle (current bounding box.north east);
        \node[draw,fill=white,ellipse,anchor=west,inner sep=1pt,minimum width=4ex] at (current bounding box.north
        west){#1};
      \end{scope}
    }}%fill=Blue!7, 
}
\tikzset{
  mybackground58/.style={execute at end picture={
        \begin{scope}[on background layer]
          \draw[black, fill=blue!40!black!5, rounded corners=1ex] (current bounding box.south west)
                    rectangle (current bounding box.north east);
          \node[draw,fill=white,ellipse,anchor=west,inner sep=1pt,minimum width=4ex, rounded corners=1ex] at (current bounding box.north
                   west){#1};
        \end{scope}
    }},
}
\tikzset{l3 switch/.style={
    parallelepiped,fill=switch, draw=white,
    minimum width=0.75cm,
    minimum height=0.75cm,
    parallelepiped offset x=1.75mm,
    parallelepiped offset y=1.25mm,
    path picture={
      \node[fill=white,
        circle,
        minimum size=6pt,
        inner sep=0pt,
        append after command={
          \pgfextra{
            \foreach \angle in {0,45,...,360}
            \draw[-latex,fill=white] (\tikzlastnode.\angle)--++(\angle:2.25mm);
          }
        }
      ]
       at ([xshift=-0.75mm,yshift=-0.5mm]path picture bounding box.center){};
    }
  },
  ports/.style={
    line width=0.3pt,
    top color=gray!20,
    bottom color=gray!80
  },
  rack switch/.style={
    parallelepiped,fill=white, draw,
    minimum width=1.25cm,
    minimum height=0.25cm,
    parallelepiped offset x=2mm,
    parallelepiped offset y=1.25mm,
    xscale=-1,
    path picture={
      \draw[top color=gray!5,bottom color=gray!40]
      (path picture bounding box.south west) rectangle
      (path picture bounding box.north east);
      \coordinate (A-west) at ([xshift=-0.2cm]path picture bounding box.west);
      \coordinate (A-center) at ($(path picture bounding box.center)!0!(path
        picture bounding box.south)$);
      \foreach \x in {0.275,0.525,0.775}{
        \draw[ports]([yshift=-0.05cm]$(A-west)!\x!(A-center)$)
          rectangle +(0.1,0.05);
        \draw[ports]([yshift=-0.125cm]$(A-west)!\x!(A-center)$)
          rectangle +(0.1,0.05);
       }
      \coordinate (A-east) at (path picture bounding box.east);
      \foreach \x in {0.085,0.21,0.335,0.455,0.635,0.755,0.875,1}{
        \draw[ports]([yshift=-0.1125cm]$(A-east)!\x!(A-center)$)
          rectangle +(0.05,0.1);
      }
    }
  },
  server/.style={
    parallelepiped,
    fill=white, draw,
    minimum width=0.35cm,
    minimum height=0.75cm,
    parallelepiped offset x=3mm,
    parallelepiped offset y=2mm,
    xscale=-1,
    path picture={
      \draw[top color=gray!5,bottom color=gray!40]
      (path picture bounding box.south west) rectangle
      (path picture bounding box.north east);
      \coordinate (A-center) at ($(path picture bounding box.center)!0!(path
        picture bounding box.south)$);
      \coordinate (A-west) at ([xshift=-0.575cm]path picture bounding box.west);
      \draw[ports]([yshift=0.1cm]$(A-west)!0!(A-center)$)
        rectangle +(0.2,0.065);
      \draw[ports]([yshift=0.01cm]$(A-west)!0.085!(A-center)$)
        rectangle +(0.15,0.05);
      \fill[black]([yshift=-0.35cm]$(A-west)!-0.1!(A-center)$)
        rectangle +(0.235,0.0175);
      \fill[black]([yshift=-0.385cm]$(A-west)!-0.1!(A-center)$)
        rectangle +(0.235,0.0175);
      \fill[black]([yshift=-0.42cm]$(A-west)!-0.1!(A-center)$)
        rectangle +(0.235,0.0175);
    }
  },
}

\usetikzlibrary{calc, shadings, shadows, shapes.arrows}
\tikzset{cross/.style={cross out, draw=black, minimum size=2*(#1-\pgflinewidth), inner sep=0pt, outer sep=0pt},
cross/.default={1pt}}
\tikzset{%
  interface/.style={draw, rectangle, rounded corners, font=\LARGE\sffamily},
  ethernet/.style={interface, fill=yellow!50},% ethernet interface
  serial/.style={interface, fill=green!70},% serial interface
  speed/.style={sloped, anchor=south, font=\large\sffamily},% line speed at edge
  route/.style={draw, shape=single arrow, single arrow head extend=4mm,
    minimum height=1.7cm, minimum width=3mm, white, fill=switch!20,
    drop shadow={opacity=.8, fill=switch}, font=\tiny}% inroute/outroute arrows
}
\newcommand*{\shift}{1.3cm}% For placing the arrows later
\newcommand*{\router}[1]{
\begin{tikzpicture}
  \coordinate (ll) at (-3,0.5);
  \coordinate (lr) at (3,0.5);
  \coordinate (ul) at (-3,2);
  \coordinate (ur) at (3,2);
  \shade [shading angle=90, left color=switch, right color=white] (ll)
    arc (-180:-60:3cm and .75cm) -- +(0,1.5) arc (-60:-180:3cm and .75cm)
    -- cycle;
  \shade [shading angle=270, right color=switch, left color=white!50] (lr)
    arc (0:-60:3cm and .75cm) -- +(0,1.5) arc (-60:0:3cm and .75cm) -- cycle;
  \draw [thick] (ll) arc (-180:0:3cm and .75cm)
    -- (ur) arc (0:-180:3cm and .75cm) -- cycle;
  \draw [thick, shade, upper left=switch, lower left=switch,
    upper right=switch, lower right=white] (ul)
    arc (-180:180:3cm and .75cm);
  \node at (0,0.5){\color{blue!60!black}\Huge #1};% The name of the router
  \begin{scope}[yshift=2cm, yscale=0.28, transform shape]
    \node[route, rotate=45, xshift=\shift] {\strut};
    \node[route, rotate=-45, xshift=-\shift] {\strut};
    \node[route, rotate=-135, xshift=\shift] {\strut};
    \node[route, rotate=135, xshift=-\shift] {\strut};
  \end{scope}
\end{tikzpicture}}

\makeatletter
\pgfdeclareradialshading[tikz@ball]{cloud}{\pgfpoint{-0.275cm}{0.4cm}}{%
  color(0cm)=(tikz@ball!75!white);
  color(0.1cm)=(tikz@ball!85!white);
  color(0.2cm)=(tikz@ball!95!white);
  color(0.7cm)=(tikz@ball!89!black);
  color(1cm)=(tikz@ball!75!black)
}
\tikzoption{cloud color}{\pgfutil@colorlet{tikz@ball}{#1}%
  \def\tikz@shading{cloud}\tikz@addmode{\tikz@mode@shadetrue}}
\makeatother
\tikzset{my cloud/.style={
     cloud, draw, aspect=2,
     cloud color={gray!5!white}
  }
}

\usetikzlibrary{backgrounds}
\usetikzlibrary{patterns}
\pgfmathdeclarefunction{gauss}{3}{%
  \pgfmathparse{1/(#3*sqrt(2*pi))*exp(-((#1-#2)^2)/(2*#3^2))}%
}
\pgfmathdeclarefunction{cdf}{3}{%
  \pgfmathparse{1/(1+exp(-0.07056*((#1-#2)/#3)^3 - 1.5976*(#1-#2)/#3))}%
}
\pgfmathdeclarefunction{fq}{3}{%
  \pgfmathparse{1/(sqrt(2*pi*#1))*exp(-(sqrt(#1)-#2/#3)^2/2)}%
}
\pgfmathdeclarefunction{fq0}{1}{%
  \pgfmathparse{1/(sqrt(2*pi*#1))*exp(-#1/2)}%
}

\usepackage{etoolbox}

\makeatletter
\tikzset{
    database/.style={
        path picture={
            \draw (0, 1.5*\database@segmentheight) circle [x radius=\database@radius,y radius=\database@aspectratio*\database@radius];
            \draw (-\database@radius, 0.5*\database@segmentheight) arc [start angle=180,end angle=360,x radius=\database@radius, y radius=\database@aspectratio*\database@radius];
            \draw (-\database@radius,-0.5*\database@segmentheight) arc [start angle=180,end angle=360,x radius=\database@radius, y radius=\database@aspectratio*\database@radius];
            \draw (-\database@radius,1.5*\database@segmentheight) -- ++(0,-3*\database@segmentheight) arc [start angle=180,end angle=360,x radius=\database@radius, y radius=\database@aspectratio*\database@radius] -- ++(0,3*\database@segmentheight);
        },
        minimum width=2*\database@radius + \pgflinewidth,
        minimum height=3*\database@segmentheight + 2*\database@aspectratio*\database@radius + \pgflinewidth,
    },
    database segment height/.store in=\database@segmentheight,
    database radius/.store in=\database@radius,
    database aspect ratio/.store in=\database@aspectratio,
    database segment height=0.1cm,
    database radius=0.25cm,
    database aspect ratio=0.35,
  }
\makeatother

\floatstyle{ruled}
\newfloat{listing}{tb}{lst}{}
\floatname{listing}{Listing}
\title{In-Context Autonomous Network Incident Response:\\ An End-to-End Large Language Model Agent Approach}
\author{
Yiran Gao\textsuperscript{\rm 1}, Kim Hammar\textsuperscript{\rm 2}, Tao Li\textsuperscript{\rm 1}\thanks{Corresponding author}
}
\affiliations{
\textsuperscript{\rm 1}Department of Systems Engineering, City University of Hong Kong, Hong Kong SAR, China\\
    \textsuperscript{\rm 2}Department of Electrical and Electronic Engineering, University of Melbourne, Australia\\
    gaoyiran525@gmail.com, kim.hammar@unimelb.edu.au, li.tao@cityu.edu.hk 
}

\begin{document}

\maketitle

\begin{abstract}
Rapidly evolving cyberattacks demand incident response systems that can autonomously learn and adapt to changing threats. Prior work has extensively explored the reinforcement learning approach, which involves learning response strategies through extensive simulation of the incident. While this approach can be effective, it requires handcrafted modeling of the simulator and suppresses useful semantics from raw system logs and alerts. To address these limitations, we propose to leverage large language models' (\textsc{llm}) pre-trained security knowledge and in-context learning to create an end-to-end agentic solution for incident response planning. Specifically, our agent integrates four functionalities, perception, reasoning, planning, and action, into one lightweight \textsc{llm} (14b model). Through fine-tuning and chain-of-thought reasoning, our \textsc{llm} agent is capable of processing system logs and inferring the underlying network state (perception), updating its conjecture of attack models (reasoning), simulating consequences under different response strategies (planning), and generating an effective response (action). By comparing \textsc{llm}-simulated outcomes with actual observations, the \textsc{llm} agent repeatedly refines its attack model and corresponding response, thereby demonstrating in-context adaptation. Our agentic approach is free of modeling and can run on commodity hardware. When evaluated on incident logs reported in the literature, our agent achieves recovery up to $23\%$ faster than those of frontier \textsc{llm}s.   
\end{abstract}

\begin{links}
    \link{Code}{https://github.com/TaoLi-NYU/llmagent4incidense-response-aaai26summer}
\end{links}

\section{Introduction}
Network incident response is a decision-making process in the post-attack stage aimed at containing, mitigating, and recovering from cyberattacks. Today's response practice relies heavily on manual operations, which can be slow and labor-intensive and fall short in the face of a continuously evolving security landscape. A recent study reports that more than 60\% of surveyed organizations take more than 100 days to recover from incidents \cite{ibm-report}. 

To address the limitations of manual responses, substantial research efforts have been dedicated to automated responses, in which response planning and execution are delegated to an artificial intelligence (\textsc{ai}) agent with minimal human intervention \cite{li2025agentic}. Prior work has extensively explored reinforcement learning (\textsc{rl}) approaches, where the incident response is formulated as a Markov decision process (\textsc{mdp}) or a game between the attack and the defense \cite{tao22info}. While \textsc{rl} agents have demonstrated success in simulations \cite{auto-defense}, their practical implementations are hindered by the stringent requirement for structured network environment modeling, which compresses semantics in system logs and security alerts into succinct numeric data. Such a practice still requires manual labor and suppresses useful semantics. 

Motivated by the emerging abilities of large language models (\textsc{llm}s), the idea of \textsc{llm} agent for autonomous defense has gained increasing momentum \cite{llm4sec-review25, li2025texts}. While this emerging direction largely falls within academic research at the moment, industry stakeholders have attempted to commercialize it, as exemplified by \textsc{ibm}'s \textsc{llm}-based incident investigation service \cite{ibm-incident-product}. Compared with \textsc{rl} agents, \textsc{llm} agents excel at handling textual data from system logs and alerts and at leveraging their built-in security knowledge when planning response actions \cite{hamoun25llm-acd, cardenas25llm4acd}, thereby sparing the manual labor of structured modeling.  

However, unlike \textsc{rl} agents that are specifically tailored to long-horizon incident response tasks, most \textsc{llm}-based methods proposed in the literature rely on prompt engineering of general-purpose \textsc{llm}s. Consequently, the \textsc{llm} agents are plagued by 1) \textbf{hallucinations}: generating response actions that appear plausible but are inappropriate, and 2) \textbf{context loss} in long-term planning: \textsc{llm}s losing track of prior context as new findings overload history information, leading to incoherent response strategies  \cite{guo25ircopilot}. 

%\tikzexternaldisable
\begin{figure*}[!ht]
  \centering
%\tikzsetnextfilename{deployment2}    
  \scalebox{1.57}{
   \input{tikz/deployment2.tex}    
  }
  \caption{Overview of the two stages of our approach. In the first stage [cf.~\textbf{a})], an \textsc{llm} is fine-tuned offline using a dataset of incident logs, each paired with corresponding response plans and chain-of-thought reasoning traces. In the second stage [cf.~\textbf{b})], the fine-tuned \textsc{llm} processes system logs and threat intelligence online to generate $N$ candidate response actions. A planning agent then evaluates these candidates through rollout and in-context adaptation, after which it selects the most effective action.}
  \label{fig:deployment}
\end{figure*}
%\tikzexternalenable
This work aims to mitigate the above shortcomings by distilling \textsc{rl}-based planning principles in partially observed \textsc{mdp} (\textsc{pomdp}) into the \textsc{llm} agentic workflows, facilitating an end-to-end incident response that directly maps logs and alerts to sequences of response actions. Our proposed agentic approach integrates four functionalities into a single lightweight \textsc{llm} with 14 billion parameters, deployable on commodity hardware. Inspired by the online lookahead rollout methods in \textsc{pomdp} \cite{tao23cola,tao24col}, the four functionalities include 1) \textbf{perception}: processing the log and security alert data and inferring the underlying network recovery state; 2) \textbf{reasoning}: combining the built-in security knowledge and conjecture of attack tactics to forecast future alerts and recovery state, which corresponds to a ``world model'' of the network environment; 3) \textbf{planning}: carrying out a lookahead tree-search by simulating different action sequences; and 4) \textbf{action}: translating the high-level response strategies to security commands.

Our \textbf{core contribution} lies in the interplay between the agent's offline fine-tuned reasoning capability and online planning to address the hallucinations and context loss, as illustrated in Fig.~\ref{fig:deployment}. The internal world model's simulated recovery trajectories will be scrutinized in the planning stage, where hallucinated actions are filtered out. Meanwhile, the planned responses will be terminated if the world model's predicted alerts deviate from the actual observations and re-planned after the \textsc{llm} calibrates the world model through in-context learning, ensuring self-consistency in long-horizon response planning. We evaluate our agent on real-world incidence log data and, our agent generates effective response plan $23\%$ faster than the frontier \textsc{llm}s and prior works.

\section{Related Work}
Towards automating security response and more broadly, autonomous cyber defense, prior and contemporary efforts have been focusing on decision/game-theoretic \cite{kmi24feedback-control,zhu13gamesec, tao24ddztd}, \textsc{rl}-based \cite{auto-defense, ge23meta-rl}, and most recently, \textsc{llm}-based approaches \cite{llm4sec-review25, li2025texts}. Compared with the other two, \textsc{llm}s can directly generate response strategies by taking in log data without mathematical modeling or extensive pre-training in simulations, thanks to their text processing, semantic understanding, and pre-trained knowledge base.   

Recent efforts on the \textsc{llm}-based methods can be categorized into two major classes: prompt-based \textsc{llm} orchestration and \textsc{llm}-\textsc{rl} hybrid agentic approaches. The first class breaks down the entire incident response into several subtasks and develops detailed prompts for \textsc{llm}s or independent \textsc{llm} sessions when tackling each task \cite{hamoun25llm-acd, guo25ircopilot, tao25deception}. Featuring end-to-end operations, such an approach requires substantial effort in designing complex prompts to reduce hallucinations and maintain prior context, preserving coherence in lengthy interactions \cite{guo25ircopilot}.

The hybrid approach alleviates these limitations by combining \textsc{rl} and \textsc{llm} agents, where \textsc{rl} agents supervise the \textsc{llm}'s generation \cite{keman24rl-mentor}, \textsc{llm} agents augment \textsc{rl} agents through knowledge sharing and human interactions \cite{lopes24hybrid-ai}, and two agents communicate with each other \cite{cardenas25llm4acd}. Despite the different nature of agentic interactions, these works require additional \textsc{rl} training in simulated environments.

Our work aims to develop an \textsc{rl}-inspired prompting for creating \textsc{llm} agents capable of handling the entire response cycle, which is less explored in the literature \cite{hammar2025incident, hongsong25rl-train-llm}. Combining the advantages of \textsc{rl} and \textsc{llm}, our proposed \textsc{llm} agent adopts an \textsc{rl}-type lookahead (rollout) planning procedure \cite{bertsekas24mpc} to address hallucinations and context loss based on the predictive analytics from \textsc{llm}'s processing of raw log and alert data.

\section{Incident Response Planning as Partially Observable Markov Decision Process}
Incident response involves restoring a network system to a secure, operational state after cyberattacks. Response planning involves analyzing attack patterns, securing forensic evidence, containing the attacker, restoring critical services, and hardening the system to prevent recurrence. Incident response, as a post-attack security mechanism, focuses on restoring service quality in a minimal \textit{response and recovery time}, during which response actions are planned and deployed. A key challenge in achieving a timely and effective response is the defender's incomplete information about the attack's scope, pattern, and severity due to limited and partial indicators of compromise, such as log files and alerts. 
To facilitate our later discussion, we present a partially observable Markov decision process (\textsc{pomdp}) to capture the nature of incomplete information in incident response; Fig.~\ref{fig:states} presents a visualization of response processes.

% \textcolor{red}{Fig.~8} presents a visual example of the \textsc{pomdp} formulation of incident response.

\paragraph{Recovery State} Throughout the discussion, we adopt a discrete time index $t\in \mathbb{N}$, with $t=0$ the start of response phase. We first introduce the \textit{recovery state} $s_t=(s^{\mathrm{c}}_t,s^{\mathrm{a}}_t,s^{\mathrm{p}}_t,s^{\mathrm{e}}_t,s^{\mathrm{h}}_t,s^{\mathrm{r}}_t)$ defined as a six-dimensional Boolean vector with each entry representing the progress in a specific stage of the response:
\begin{enumerate}[label=\arabic*), nosep]
    \item \textbf{Containment}: $s^{\mathrm{c}}_t$ indicates whether the attack has been isolated and prevented from spreading.
    \item \textbf{Assessment}: $s^{\mathrm{a}}_t$ indicates if the scope and severity of the attack have been identified.
    \item \textbf{Preservation}: $s^{\mathrm{p}}_t$ indicates if the forensic evidence related to the incidence has been preserved.
    \item \textbf{Eviction}: $s^{\mathrm{e}}_t$ indicates whether the attacker's access has been revoked and malicious processes terminated.
    \item \textbf{Hardening}: $s^{\mathrm{h}}_t$ indicates if the system has been hardened to prevent future recurrence.
    \item \textbf{Restoration}: $s^{\mathrm{r}}_t$ indicates whether services and user access have been restored 
\end{enumerate}

%\tikzexternaldisable
\begin{figure}
  \centering
  \scalebox{1.35}{
%\tikzsetnextfilename{states2} 
%   \includegraphics{states2.pdf}
   \begin{tikzpicture}

\node[draw,circle, minimum width=10mm, scale=0.45, fill=Blue!10](s0) at (0,0) {\large $s_0$};
\node[draw,circle, minimum width=10mm, scale=0.45](s1) at (0,-0.75) {\small $s_1,s_2$};
\node[draw,circle, minimum width=10mm, scale=0.45](s2) at (-1,-1.5) {\large $s_3$};
\node[draw,circle, minimum width=10mm, scale=0.45](s3) at (1,-1.5) {\large $s'_3$};
\node[draw,circle, minimum width=10mm, scale=0.45](s4) at (0,-2.25) {\large $s_4$};
\node[draw,circle, minimum width=10mm, scale=0.45](s5) at (0,-3) {\large $s_5$};
\node[draw,circle, minimum width=10mm, scale=0.45](s6) at (0,-3.75) {\large $s_6$};
\node[draw,circle, minimum width=10mm, scale=0.45, fill=Red!20](s8) at (0,-4.5) {\large $s_{\tau}$};

\draw[-{Latex[length=1.3mm]}, line width=0.17mm] (s0) to (s1);
\draw[-{Latex[length=1.3mm]}, line width=0.17mm] (s1) to (s2);
\draw[-{Latex[length=1.3mm]}, line width=0.17mm] (s1) to (s3);
\draw[-{Latex[length=1.3mm]}, line width=0.17mm, looseness=5, out=110, in=170] (s1) to (s1);
\draw[-{Latex[length=1.3mm]}, line width=0.17mm] (s3) to (s4);
\draw[-{Latex[length=1.3mm]}, line width=0.17mm] (s2) to (s4);
\draw[-{Latex[length=1.3mm]}, line width=0.17mm] (s4) to (s5);
\draw[-{Latex[length=1.3mm]}, line width=0.17mm] (s5) to (s6);
\draw[-{Latex[length=1.3mm]}, line width=0.17mm] (s6) to (s8);
\draw[-{Latex[length=1.3mm]}, line width=0.17mm, bend right=35] (s5) to (s8);

\node[inner sep=0pt,align=center, scale=0.55, color=black] (hacker) at (2,-0.75) {
\textsc{isolate}
};
\node[inner sep=0pt,align=center, scale=0.55, color=black] (hacker) at (2,0) {
\textsc{initiate}
};
\node[inner sep=0pt,align=center, scale=0.55, color=black] (hacker) at (2.06,-1.5) {
\textsc{assess}
};
\node[inner sep=0pt,align=center, scale=0.55, color=black] (hacker) at (1.95,-2.25) {
\textsc{preserve}
};
\node[inner sep=0pt,align=center, scale=0.55, color=black] (hacker) at (2.13,-3) {
\textsc{evict}
};
\node[inner sep=0pt,align=center, scale=0.55, color=black] (hacker) at (2,-3.75) {
\textsc{harden}
};
\node[inner sep=0pt,align=center, scale=0.55, color=black] (hacker) at (2,-4.5) {
\textsc{restore}
};
%\node[inner sep=0pt,align=center, scale=0.5, color=black] (hacker) at (2.8,0.3) {
%$\mathbf{s} = (s^{\mathrm{I}}, s^{\mathrm{S}}, s^{\mathrm{F}}, s^{\mathrm{E}}, s^{\mathrm{H}},  s^{\mathrm{R}})$
%};
\node[inner sep=0pt,align=center, scale=0.5, color=black] (hacker) at (2.55,0.22) {
$s^{\mathrm{c}}$
};
\node[inner sep=0pt,align=center, scale=0.5, color=black] (hacker) at (2.75,0.22) {
$s^{\mathrm{a}}$
};
\node[inner sep=0pt,align=center, scale=0.5, color=black] (hacker) at (2.94,0.22) {
$s^{\mathrm{p}}$
};
\node[inner sep=0pt,align=center, scale=0.5, color=black] (hacker) at (3.12,0.22) {
$s^{\mathrm{e}}$
};
\node[inner sep=0pt,align=center, scale=0.5, color=black] (hacker) at (3.31,0.22) {
$s^{\mathrm{h}}$
};
\node[inner sep=0pt,align=center, scale=0.5, color=black] (hacker) at (3.51,0.22) {
$s^{\mathrm{r}}$
};

\node[inner sep=0pt,align=center, scale=0.55, color=black] (hacker) at (3,-0.75) {
$(\text{\textcolor{Red}{$1$}},0,0,0,0,0)$
};
\node[inner sep=0pt,align=center, scale=0.55, color=black] (hacker) at (3,0) {
$(0,0,0,0,0,0)$
};
\node[inner sep=0pt,align=center, scale=0.55, color=black] (hacker) at (3,-1.5) {
$(1,\text{\textcolor{Red}{$1$}},0,0,0,0)$
};
\node[inner sep=0pt,align=center, scale=0.55, color=black] (hacker) at (3,-2.25) {
$(1,1,\text{\textcolor{Red}{$1$}},0,0,0)$
};
\node[inner sep=0pt,align=center, scale=0.55, color=black] (hacker) at (3,-3) {
$(1,1,1,\text{\textcolor{Red}{$1$}},0,0)$
};
\node[inner sep=0pt,align=center, scale=0.55, color=black] (hacker) at (3,-3.75) {
$(1,1,1,1,\text{\textcolor{Red}{$1$}},0)$
};
\node[inner sep=0pt,align=center, scale=0.55, color=black] (hacker) at (3,-4.5) {
$(1,1,1,1,1,\text{\textcolor{Red}{$1$}})$
};

\node[inner sep=0pt,align=center, scale=0.45, color=black] (hacker) at (0,0.42) {
\textit{logs indicates anomalous}\\\textit{activity on a host}
};
\node[inner sep=0pt,align=center, scale=0.45, color=black] (hacker) at (0.78,-0.4) {
$a_0$: \textit{segment network}\\
\textit{to isolate host}
};

\node[inner sep=0pt,align=center, scale=0.45, color=black] (hacker) at (-0.91,-0.5) {
\textcolor{black}{$a_1$: \textit{malware}}\\
\textit{scan \textcolor{Red}{(failed)}}
};

\node[inner sep=0pt,align=center, scale=0.45, color=black] (hacker) at (-1.43,-1) {
\textcolor{black}{$a_2$: \textit{analyze logs}}\\
\textcolor{Blue}{\textit{(found malicious process)}}
};
\node[inner sep=0pt,align=center, scale=0.45, color=black] (hacker) at (1.65,-1.08) {
\textcolor{black}{$a'_2$: \textit{analyze processes}}\\
\textcolor{Blue}{\textit{(found malicious process)}}
};

\node[inner sep=0pt,align=center, scale=0.45, color=black] (hacker) at (0,-1.6) {
$a_3$: \textit{memory dump}
};
\node[inner sep=0pt,align=center, scale=0.45, color=black] (hacker) at (0.68,-2.65) {
$a_4$: \textit{stop process}
};

\node[inner sep=0pt,align=center, scale=0.45, color=black] (hacker) at (0.7,-3.35) {
$a_5$: \textit{upgrade}\\
\textit{affected software}
};

\node[inner sep=0pt,align=center, scale=0.45, color=black] (hacker) at (-0.84,-3.75) {
$a'_5$: \textit{live-patch}\\
\textit{vulnerability}
};

\node[inner sep=0pt,align=center, scale=0.45, color=black] (hacker) at (0.72,-4.15) {
$a_6$: \textit{restart service}
};

%\node[inner sep=0pt,align=center, scale=0.55, color=black] (hacker) at (3,0) {
%\textsc{initiation} $(0,0,0,0,0,0)$
%};
%\node[inner sep=0pt,align=center, scale=0.55, color=black] (hacker) at (3,-2) {
%\textsc{assessment} $(1,1,0,0,0,0)$
%};
%\node[inner sep=0pt,align=center, scale=0.55, color=black] (hacker) at (3,-3) {
%\textsc{preservation} $(1,1,1,0,0,0)$
%};
%\node[inner sep=0pt,align=center, scale=0.55, color=black] (hacker) at (3,-4) {
%\textsc{eviction} $(1,1,1,1,0,0)$
%};
%\node[inner sep=0pt,align=center, scale=0.55, color=black] (hacker) at (3,-5) {
%\textsc{hardening} $(1,1,1,1,1,0)$
%};
%
%\node[inner sep=0pt,align=center, scale=0.55, color=black] (hacker) at (3,-6) {
%\textsc{recovery} $(1,1,1,1,1,1)$
%};

%\node[inner sep=0pt,align=center, scale=0.55, color=black] (hacker) at (0.1,-2.1) {
%$(0,0,0,0,0,0)$
%};
%\node[inner sep=0pt,align=center, scale=0.37, color=black] (hacker) at (0.1,-2.3) {
%$s^{\mathrm{I}}_t, s^{\mathrm{S}}_t, s^{\mathrm{F}}_t, s^{\mathrm{E}}_t, s^{\mathrm{H}}_t,  s^{\mathrm{R}}_t$
%};

\end{tikzpicture}        
  }
  \caption{Two example evolutions of the recovery state $s_t$. The first recovery trajectory involves the actions $a_0,a_1,a_2,a_3,a_4, a_5, a_6$ and the second trajectory involves the actions $a_0, a_1, a^{\prime}_2, a_3, a_4, a^{\prime}_5$.}
  \label{fig:states}
\end{figure}
%\tikzexternalenable

\paragraph{Partial Observation} Due to the network's complex nature and advanced attack methods, the defender is uncertain about the actual recovery state, e.g., if the attack's scope is correctly identified, properly contained, and successfully removed. It relies on system logs and security alerts from an intrusion detection device to infer the Boolean vector. We denote by $o_t$ these textual data related to the defender's partial information, see Fig.~\ref{fig:incidence-example} for an example alert. 

\paragraph{State Transition} Based on its understanding to the network environment, the defense agent plans a sequence of recovery actions $a_t, a_{t+1, \ldots}$ to be executed, and each action $a_t$ will influence the next recovery sate $s_{t+1}$, which is captured by the transition kernel $P_{\theta}(s_{t+1}|s_{t}, a_t)$, where $\theta$ encapsulate the attack's influence on the future recovery state. In practice, $\theta$ usually corresponds to attack tactics, techniques, and procedures (\textsc{ttp}) as recorded in the \textsc{miter att\&ck} knowledge base \cite{mitre-attack}. Since the partial observation also depends on the attack \textsc{ttp}, following the convention in \textsc{pomdp} studies, we incorporate the observation kernel into the state transition: $(s_{t+1}, o_{t+1})\sim P_\theta(\cdot|s_t,a_t)$.

\paragraph{Recovery Time} Since the timely response is the primary concern of our study, we associate each pair of state and recovery action with a time cost $c(s_t,a_t)$, representing the required time (in minutes) to execute the action in a certain state. For instance, the time to isolate a compromised host may be a few seconds, whereas the time to perform a system scan and configuration updates of affected systems can be around half an hour. The total recovery time $J$ is given by the cumulative time to reach the terminal state $s_T=(1,1,1,1,1,1)$, i.e., $J(s_0)\triangleq\sum_{t=0}^{\tau-1} c(s_t, a_t)$, $s_T\sim P_\theta(\cdot|s_{\tau-1}, a_{\tau-1})$ with probability 1. 

In summary, we aim to find a response policy generated by the \textsc{llm} agent, denoted by $\Phi(\cdot)$, such that the total recovery time is minimized:
    \begin{align}
    &\min\; J(s_0)=\sum_{t=0}^{\tau-1} c(s_t, a_t) \label{eq:pomdp}\\
    &\text{s.t.}\;  a_t=\Phi(o_{0:t-1}, a_{0:t-1}); (s_{t}, o_t)\sim P_\theta(\cdot|s_{t-1},a_{t-1}). \nonumber
\end{align}

\section{\textsc{llm} Agent for In-Context Adaptive Response}
\paragraph{Perception} The \textsc{llm} agent maintains an estimate of the underlying recovery state, denoted by $\hat{s}_t$, based on past observations and actions: $\hat{s}_t\sim \Phi(o_{0:t-1}, a_{0:t-1})$. For simplicity, we denote by $h_{t-1}\triangleq \{o_{0:t-1}, a_{0: t-1}\}$ the history information. To steer the \textsc{llm} agent toward effective estimation, we fine-tune it using supervised learning on a dataset of $50,\,000$ instruction-answer pairs $\mathcal{D}=\{(\mathbf{x}^i, \mathbf{y}^i)\}_{i=1}^K$, where each instruction $\mathbf{x}^i$ consists of information related to an incident and an instruction to assess the current recovery state; see Fig.~\ref{fig:incidence-example} for an incidence example. The associated answer $\mathbf{y}^i$ describes the ground-truth, which is tokenized into $\mathbf{y}^i=(y^i_1, \ldots, y^i_{\ell_i})$, following the world-level tokenization with $\ell_i$ being the token length. We pair each data point with a sequence of chain-of-thought (\textsc{cot}) reasoning steps to explain the answers \cite{dale22cot}.

Given the training dataset $\mathcal{D}$, the fine-tuning proceeds by iteratively sampling a batch of instruction-answer pairs $(\mathbf{x}^1, \mathbf{y}^1), \ldots, (\mathbf{x}^B, \mathbf{y}^B)$ and updating the \textsc{llm}'s parameters via gradient descent on the cross-entropy loss
\begin{equation}
    L(w)=-\frac{1}{B}\sum_{i=1}^B\sum_{k=1}^{\ell_i} \log \Phi_{w}(y_k^i|\mathbf{x}^i, y_{1:k-1}^{i
    }),
\end{equation}
where $w$ denotes the trainable parameters in Low-Rank Adaptation (LoRA), a widely used parameter-efficient fine-tuning technique \cite{hu2022lora}. 
\begin{figure}[h]
\begin{tcolorbox}[
  title=Incident Example from CTU-Malware-2014,
  colback=gray!5,
  colframe=gray!60,
  boxrule=0.5pt,
  arc=2pt,
  left=4pt,
  right=4pt,
  top=4pt,
  bottom=4pt
]
\small
\textbf{System description}: Two subnetworks (A and B) are connected via a switch that is also connected to the Internet. All servers run \textsc{windows xp sp2}. Their IPs are...

\textbf{Snort alert logs}: 
\begin{verbatim}
[120:3:2](http_inspect) NO CONTENT-LENGTH..
[1:31033:6]MALWARE Win.Trojan.Cryptodefence
[Classification:A Network Trojan Detected] 
[Priority 1]
{TCP}147.32.84.165:1057->222.88.205.195:443 
\end{verbatim}

\textbf{Incident description}: Server 147.32.84.165 is infected with the \textsc{win.trojan.cryptodefense} ransomware. Alerts show the server
is making outbound command and control (\textsc{c}2) connections to
222.88.205.195. ...

\textbf{Response actions}:

1. Disconnect the Ethernet cable of the infected server at
147.32.84.165 to sever its network connection. Concurrently, configure a rule on the main switch/firewall to block all outbound
traffic to the \textsc{c}2 server 222.88.205.195.

2. Analyze the central switch to scan all network traffic from
both subnetworks A and B for any other hosts attempting to make
connections to the malicious IP 222.88.205.195.

3. Before altering the infected server, create a complete bit-for-bit
forensic image of its hard drive. 

4. Wipe the hard drive of 147.32.84.165. If other infected machines
were discovered, they must also be taken offline and wiped.

5. Upgrade all servers from \textsc{windows xp sp2}
to a modern operating system that receives security patches.

6. Restore the server’s data from a trusted backup. Once the server is
rebuilt with a modern operating system, reconnect it to the network
and closely monitor for any anomalous activity.
\end{tcolorbox}

\caption{An incidence example from \cite{GARCIA2014100}.}
    \label{fig:incidence-example}
\end{figure}

\paragraph{Reasoning}
% attack alerts prediction, and in-context learning
We aim to enable the \textsc{llm} agent to predict future observations based on its understanding of the underlying state and attack assessment, which are continuously calibrated as new observations emerge. Specifically, the \textsc{llm} agent first generates observation (alert) prediction $\hat{o}_t$ based on past observations $o_{o:t-1}$ and state estimate $\hat{s}_t$, $\hat{o}_t\sim \Phi(h_{t-1}, \hat{s}_t)$; See Fig.~\ref{fig:state-prompt} and \ref{fig:alert-prompt} for the generation prompts. By combining the state estimate and future observation prediction, the agent essentially creates an internal ``world model'' of the network environment and the attack, which helps simulate the possible consequences of different response actions. Given a sequence of actions to be evaluated through simulation starting from time $t$: $\{a_t, a_{t+1}, \ldots\}$, the \textsc{llm} agent can simulate a \textit{recovery trajectory} as below. For $\tau=t, t+1, \ldots$,
\begin{align*}
    \hat{s}_{\tau+1}\sim \Phi(h_{t}, \hat{o}_{t:\tau}, a_{t:\tau}),\quad \hat{o}_{\tau+1}\sim \Phi(h_t, \hat{o}_{t:\tau}, a_{t:\tau}, \hat{s}_{\tau+1} ),
\end{align*}
which essentially creates a world model $P_{\Phi}(\hat{s}_{t+1}|\hat{s}_t, a_t)$. The fine-tuning of the observation (alert) generation follows the same practice as in the perception part, where the answers become the alert classification and priority as shown in Fig.~\ref{fig:alert-prompt}. Note that the attack tactic is required in the fine-tuning when predicting alerts. As discussed in the following paragraph, the agent needs to consider possible tactics during the planning stage. 
\begin{figure}[h]
\begin{tcolorbox}[title=State Generation Prompt Template,
  colback=bluetwo!20,
  colframe=bluetwo,
  boxrule=0.5pt,
  arc=2pt,
  left=4pt,
  right=4pt,
  top=4pt,
  bottom=4pt]
    \small
    \textbf{\#\#\#System description}: ...

    \textbf{\#\#\#logs}: ...

    \textbf{\#\#\#Incident description}:...

    \textbf{\#\#\#\textsc{miter att\&ck} tactics being used}: Initial Access, Execution, Persistence.

    \textbf{\#\#\#\textsc{miter att\&ck} techniques being used}: T1190 Exploit Public-Facing Application.

    \textbf{\#\#\# Previous State}: ``is\_attack\_contained'' : true; ``is\_knowledge\_sufficient'': false,
    ``are\_forensics\_preserved'': false,
    ``is\_eradicated'': false,
    ``is\_hardened'': false,
    ``is\_recovered'': false.

\textbf{\#\#\#Previous recovery actions}: 

Action: Acquire full disk and memory images of REDIS01, collect Redis RDB/AOF files, NAT/firewall logs, and export system logs to write-protected storage. 

Explanation: Preserves evidence and provides the data needed for thorough eradication.

\textbf{\#\#\#Instruction}: 
You have been given information about a security incident, the state of recovery from the incident, and a recovery action.
Your task is to predict what the next state of the recovery will be after applying the recovery action.
For example, if the given recovery action effectively contains the attack and ``is\_attack\_contained'' is ``false'' in the current state, then the next state should have ``is\_attack\_contained'' set to ``true''.
It is also possible that multiple state properties change values from false to true. It is also possible that the state remains the same, i.e., no property changes.
It is important that the state only changes if the action is effective in achieving one of the recovery goals: containment, information gathering, preserving evidence, eradication, hardening, or recovery.
A state variable can only change from ``false'' to ``true'', it cannot be changed from ``true'' to ``false''.

\textbf{\#\#\# Response}:
<think>

\end{tcolorbox}
    \caption{The prompt for generating recovery states. The instruction presents an example of recovery state transition under attack tactics/techniques and recovery actions. }
    \label{fig:state-prompt}
\end{figure}
\begin{figure}[ht]
\begin{tcolorbox}[title=Alert Generation Prompt Template,
  colback=orange!10,
  colframe=orange!60,
  boxrule=0.5pt,
  arc=2pt,
  left=4pt,
  right=4pt,
  top=4pt,
  bottom=4pt]
    \small
    \textbf{\#\#\#System description}: ...

    \textbf{\#\#\#logs}: ...

    \textbf{\#\#\#Incident description}:...

    \textbf{\#\#\#\textsc{miter att\&ck} tactics being used}: Initial Access, Execution, Persistence.

    \textbf{\#\#\#\textsc{miter att\&ck} techniques being used}: T1190 Exploit Public-Facing Application.

    \textbf{\#\#\# Instruction}:
Generate fields produced by an intrusion detection system (e.g., Snort) during a cyberattack by an attacker following this \textsc{mitre att\&ck} tactic: Impact. Frame your answers as ``[Classification: alert type] [Priority: level].''

\textbf{\#\#\# Response}:
<think>
    \end{tcolorbox}
    \caption{The prompt for generating alerts. The \textsc{llm} generations are structured into pairs of alert-type classifications, e.g., Bad Traffic and Network Trojans, and priority levels. }
    \label{fig:alert-prompt}
\end{figure}
\vspace{-1em}
\begin{figure}[ht]
\begin{tcolorbox}[title=Attack Tactics Calibration Prompt Template,
  colback=RedOrange!10,
  colframe=RedOrange!70,
  boxrule=0.5pt,
  arc=2pt,
  left=4pt,
  right=4pt,
  top=4pt,
  bottom=4pt]
    \small
    \textbf{\#\#\#System description}: ...

    \textbf{\#\#\#logs}: ...

    \textbf{\#\#\#Incident description}:...

    \textbf{\#\#\#\textsc{miter att\&ck} tactics being used}: Initial Access, Execution, Persistence.

    \textbf{\#\#\#\textsc{Predicted alerts characteristics}}: ``[Classification: alert type] [Priority: level].''

    \textbf{\#\#\#\textsc{Observed alerts characteristics}}: ``[Classification: alert type] [Priority: level].''

    \textbf{\#\#\# Instruction}:
Your task is to reassess the previously chosen MITRE ATT\&CK tactic label(s), using the evidence in the above fields and comparing the predicted and observed alert characteristics. You must propose revised tactic candidates STRICTLY from the provided candidate tactics set, which includes the common tactics for such an incidence. 

\textbf{\#\#\# Response}:
<think>
    \end{tcolorbox}    
   
    \caption{The prompt for attack tactics conjecture calibration using \textsc{gpt}-5.2. }
    \label{fig:calibration}
\end{figure}

\begin{figure}[h]
\begin{tcolorbox}[title=Action Generation Prompt Template,
  colback=teal!5,
  colframe=teal!60,
  boxrule=0.5pt,
  arc=2pt,
  left=4pt,
  right=4pt,
  top=4pt,
  bottom=4pt]
    \small
    \textbf{\#\#\#\textsc{miter att\&ck} tactics being used}: Initial Access, Execution, Persistence.

    \textbf{\#\#\#\textsc{miter att\&ck} techniques being used}: T1190 Exploit Public-Facing Application.

    \textbf{\#\#\# Previous State}:...

\textbf{\#\#\#Previous recovery actions}: 

Action: Acquire full disk and memory images of REDIS01, collect Redis RDB/AOF files, NAT/firewall logs. 

Explanation: Preserves evidence and provides the data needed for thorough eradication.

\textbf{\#\#\#Instruction}: You are a security operator ... . The goal when selecting the recovery action is to change the state so that one of the state-properties that is currently ``false'' becomes ``true''. The ideal recovery action sequence is: 1. contain the attack 2. gather information 3. preserve evidence 4. eradicate the attacker 5. harden the system 6. recover operational services.
When selecting the recovery action, make sure that it is concrete and actionable and minimizes unnecessary service disruptions. Vague or unnecessary actions will not change the state and should be avoided.
Return a JSON object with two properties: ``Action'' and ``Explanation'', both of which should be strings.
The property ``Action'' should be a string that concisely describes the concrete recovery action.
The property ``Explanation'' should be a string that concisely explains why you selected the recovery action and motivates why the action is needed.

\textbf{\#\#\# Response}:
<think>
    \end{tcolorbox}
    \caption{The prompt for generation recovery actions. The \textsc{llm} is instructed to follow an ideal recovery trajectory and explain its motivations.}
    \label{fig:action-prompt}
\end{figure}

\paragraph{Planning} 
% conjectural learning
Utilizing the \textsc{llm}-based cyber world model, we propose an online conjectural lookahead planning method, inspired by an \textsc{rl} paradigm: Monte-Carlo tree search under misspecification in \textsc{pomdp} \cite{kim-tao25col, kim-tao25quantization}. The intuition behind such an \textsc{rl} algorithm is to first predict future consequences under diverse actions using the world model, which may be inaccurate (mispecified with respect to the true model), and then execute the best-performing one (under misspecification). The gap between newly emerged alerts and the predicted outcomes from the world model will provide context for \textsc{llm} to reflect on its early conjectures and refine its action generation. 

Specifically, the \textsc{llm} agent starts by conjecturing attack tactics by screening system logs and incident information, then selecting a tactic from a set of plausible tactics, denoted by $\hat{\theta}\in \Theta$. At each time of the response, we prompt the agent to 1) infer the underlying recovery state $\hat{s}_t$; 2) generate $N$ candidate actions $\mathcal{A}_t=\{\hat{a}_t^1, \hat{a}_t^2,\ldots, \hat{a}_t^N\}$, and 3) for each action, predict the next recovery state and alert, $(\hat{s}_{t+1}, \hat{o}_{t+1})\sim \Phi(h_t, \hat{a}_{t}^k)$, $k=\in \{1, \ldots, N\}\triangleq [N]$, leveraging perception and reasoning functionalities.  Given the generated, tentative actions, the agent needs to identify the best performer by 4) simulating $M$ recovery trajectories starting from $(\hat{s}_{t+1}, \hat{a}_{t+1}^k)$, which is defined as a sequence of state-action pair generated by the \textsc{llm} until the terminal sate $\hat{s}_T=(1,1,1,1,1,1)$: $q^i\triangleq\{(\hat{s}_{t+1}, \hat{a}_{t+1}^k), (\hat{s}_{t+2}^{i}, \hat{a}_{t+1}^{k,i}), \ldots, \hat{s}_T\}$, $i\in \{1, \ldots, M\}\triangleq[M]$. 5) The estimated cumulated cost associated with $\hat{a}_{t+1}^k$ is given by the sample average, denoted by the Q function, based on which the agent selects the cost minimizer:
\begin{subequations}
  \begin{align}
    & Q(\hat{s}_{t+1}, \hat{a}_{t+1}^k)=\frac{1}{M} \sum_{i\in [M]} \sum_{(\hat{s}, \hat{a})\in q^i}c(\hat{s},\hat{a}),\\
    & a_{t+1}\in \argmin_{a\in \mathcal{A}_t} Q(\hat{s}_{t+1}, a).
\end{align}  
\end{subequations}

The executed action incurs actual alerts $o_{t+1}\sim P_{\theta}(s_{t+1}, a_{t+1})$, which enables the agent to 6) compare the actual and predicted alerts and calibrate the conjectured attack tactics if there exists a significant discrepancy between the two. We rely on a frontier \textsc{llm}, \textsc{gpt}-5.2, to digest the newly emerged alert information and recommend a more likely alternative: $\hat{\theta}_{t+1}\leftarrow \textsc{gpt}(\hat{o}_{t+1}, o_{t+1}, a_{t+1})$, see Fig.~\ref{fig:calibration} for the calibration prompt. Note that such a calibration only requires \textsc{api} access to the frontier models without local deployment, which remains commodity hardware-friendly. Moreover, such calibration can also be performed by the proposed \textsc{llm} agent itself after retrieving relevant information from external threat intelligence, which is one of the future extensions. Finally, the agent repeats the same planning procedure under the calibrated conjecture from $t+1$. Algo.~\ref{algo:conjecture-plan} summarizes the planning cycle.

\begin{algorithm}[h]
\caption{Online Conjectural Lookahead Planning with In-Context Adaptation}
\label{algo:conjecture-plan}
    \begin{algorithmic}[1]
        \STATE \textbf{Input:} \textsc{llm} $\Phi$, system logs, action batch $N$, sample trajectory batch $M$. 
        \STATE \textbf{Output:} Response actions $\pi=\{a_1, \ldots, a_T\}$.
        \STATE \textbf{Initialization} $\hat{s}_0\leftarrow (0,0,0,0,0,0)$, $\hat{\theta}_0$, $\pi\leftarrow\{\}$, $t\leftarrow 0$.
        \WHILE{$\hat{s}_t\neq (1,1,1,1,1,1)$}
        \STATE Sample $\{\hat{a}_t^1, \hat{a}_t^2, \ldots, \hat{a}_t^N\}=\mathcal{A}_t$ from $\Phi(\cdot|h_t)$.
        \FOR{$i\in [M]$}
        \STATE $Q(\hat{s}_t, \hat{a}_t^i)=\frac{1}{M} \sum_{k=1}^M \textsc{recovery-to-go}(\hat{s}_t, \hat{a}_t^i)$
        \ENDFOR
        \STATE Execute $a_t=\argmin_{a\in \mathcal{A}_t} Q(\hat{s}_t, \hat{a}_t^i)$.
        \STATE Simulate $(\hat{s}_{t+1}, \hat{o}_{t+1})\sim \Phi(h_t, a_t)$, and receive the alert $o_{t+1}$.
        \STATE Context adaptation $\hat{\theta}_{t+1}\gets \textsc{gpt}(\hat{o}_{t+1}, o_{t+1}, a_{t+1})$.
        \STATE $h_{t+1}\gets h_t\cup\{o_{t+1}, a_t\}$, $t\gets t+1$. 
        \ENDWHILE
        \PROCEDURE{recovery-to-go}{s,a}
        \STATE Simulate $s'\sim \Phi(s,a)$.
        \IF{ $s'=(1,1,1,1,1,1)$}
        \RETURN c(s,a)
        \ENDIF
        \STATE Sample $a'\sim \Phi(\cdot|s')$.
        \RETURN $c(s,a)+\textsc{recovery-to-go}(s',a')$.
        \ENDPROCEDURE
    \end{algorithmic}
\end{algorithm}

\paragraph{Action} During the planning stage, the \textsc{llm} agent needs to generate candidate actions based on its perception of the current recovery status. Towards this end, we fine-tune the \textsc{llm} to generate effective response actions with reduced hallucinations and consistent with previous actions. Similar to state- and alert-generation fine-tuning, we present the \textsc{llm} with pairs of instructions and answers, adding the previous estimated recovery state, previous recovery actions, and conjectured attack tactics to the instruction. Fig.~\ref{fig:action-prompt} presents the prompt template. For actions that lead to longer recovery paths and for predicted alerts that are inconsistent with actual ones, we regard them as likely instances of hallucinations and context-rot generation and will filter them out.   

\section{Experiment}

\subsection{Perception \& Reasoning: LoRA Fine-Tuning }
\label{subsec:lora_sft_style}

\paragraph{Experiment Setup}
We fine-tune a DeepSeek-14B (Qwen-compatible) language model using LoRA-based supervised fine-tuning on \texttt{CSLE-IncidentResponse-V1} (\texttt{states\_examples.json}) \cite{kim24dataset}.  We take the first $50,\,000$ instruction--answer pairs from the dataset and train LoRA adapters with the hyperparameters detailed in Table~\ref{tab:lora}.
\begin{table}[!ht]
    \centering
    \caption{A summary of hyperparameters}
    \begin{tabular}{ll}
    \toprule
    Parameter(s)     &   Value(s)\\
    \midrule
    LoRA rank, scaling, dropout & 64,128,0.05\\ learning rate & 0.00095 \\ {per\_device\_batch\_size} & 2\\ {gradient\_accumulation\_steps} & 16 \\ temperature & 0.6 \\
    \midrule
    action generation batch & 3\\
    recovery trajectory batch & 3\\
    \bottomrule
    \end{tabular}
    
    \label{tab:lora}
\end{table}

\paragraph{Evaluation Setup and Metrics}
The evaluation data are randomly sampled, excluding the training data, and we collect a total of $17,\,600$ pairs for the testing dataset. We use two metrics to measure the prediction performance. 1) {Exact-match accuracy}: We treat both the prediction and the ground-truth label as JSON text and count a prediction as correct only if it matches the label \emph{exactly} on all entries. This metric is intentionally strict and captures whether the model can produce a fully correct, structured output in a single shot. The exact accuracy is $0.98$.
2) {Multi-label F1}: since recovery-state prediction consists of multiple Boolean entries, it is naturally a multi-label binary classification problem. We therefore
report two averaged F1 scores:
(i) {class-agnostic average F1} (caa-F1), which aggregates True Positive (TP)/ False Positive (FP)/ False Negative (FN) across
all entries before computing F1, and (ii) {class-specific average F1} (csa-F1),
which computes F1 per Boolean entry and then averages across entries. In addition, we report {Per-entry F1} to diagnose which entries are harder to predict or exhibit systematic bias. Table~\ref{tab:f1} summarizes the F1 scores, which suggests that our model can accurately estimate the underlying state.

\paragraph{Alert Prediction Evaluation}
We further evaluate the model’s ability to predict intrusion-detection system (\textsc{ids}) alert fields
(\texttt{classification} \& \texttt{priority}) using the prompt shown in Fig.~\ref{fig:alert-prompt}. We adopt a \emph{unique-pair precision/recall} metric detailed below.

Given one evaluation sample with instruction prompt $\mathbf{x}$, we obtain the model's prediction $\hat{\mathbf{y}}\sim \Phi(\cdot|\mathbf{x})$ to be compared with the ground-truth $\mathbf{y}$. Since $\hat{\mathbf{y}}$ contains the full prompt followed by the generated continuation, we remove the prompt
prefix to keep only the generated part. We then extract all alert tuples of the form
\([\texttt{Classification: }\ldots]\,[\texttt{Priority: }\ldots]\)
from both $\hat{\mathbf{y}}$ and $\mathbf{y}$ using a regular expression, and remove duplicates to obtain the
\emph{unique} prediction set $\widehat{\mathcal{P}}$ and label set $\mathcal{P}$. For each sample, we compute the overlap $|\widehat{\mathcal{P}} \cap \mathcal{P}|$ and define $\mathrm{Precision} \triangleq |\widehat{\mathcal{P}} \cap \mathcal{P}|/ |\widehat{\mathcal{P}}|$, $\mathrm{Recall}\triangleq |\widehat{\mathcal{P}} \cap \mathcal{P}|/|\mathcal{P}|$.

We report the F1 score for alert generation in Table~\ref{tab:label_f1} across the top 5 most frequent tactics, from which we find that our model is better suited to handle generation under attack than in normal activity, since the corresponding alerts are false alarms and do not exhibit any pattern. 
\begin{table}[]
    \centering
    \resizebox{\linewidth}{!}{
    \begin{tabular}{lllllllll}
    \toprule
            &  caa & csa & $s^c$ & $s^a$ & $s^p$& $s^e$ & $s^h$ & $s^r$  \\
    \midrule
       F1  & 0.9902 & 0.9822 & 0.9975 & 0.9964 & 0.9970& 0.9952 & 0.9541 & 0.9533\\
    \bottomrule
    \end{tabular}
    }
    \caption{A summary of F1 scores under class-agnostic average (caa) and class-specific average (csa).}
    \label{tab:f1}
\end{table}

\begin{table}[t]
\centering
\small
\resizebox{\linewidth}{!}{
\begin{tabular}{ll}
\toprule
\textbf{Tactics}(\% over all data points)  & \textbf{F1} \\
\hline
{\small Normal Activity (15.59\%)}  & 0.5711 \\
{\small Initial Access, Execution, Collection, Exfiltration (6.92\%)}  & 0.8579 \\
{\small  Access, Execution, Credential Access, Exfiltration (1.71\%)}   & 0.8599 \\
{\small Impact  (1.55\%)} & 0.8758 \\
{\small  Access, Execution, Command and Control, Exfiltration (1.52\%)}   & 0.7424 \\
\bottomrule
\end{tabular}
}
\caption{Most frequent tactics and corresponding F1 scores.}
\label{tab:label_f1}
\end{table}

\begin{table*}[!ht]
    \centering
    \begin{tabular}{llll}
    \toprule
    Dataset & Systems & Attacks & logs\\
    \midrule
    CTU-Malware-2014     & \textsc{windows xp sp2} & Malwares and ransomwares & \textsc{snort} alerts\\
    CIC-IDS-2017     & \textsc{windows}, \textsc{linux} & Denial-of-Service, web attacks, \textsc{sql} injections, and etc.&\textsc{snort} alerts\\
    AIT-IDS-V2-2022 & \textsc{windows}, \textsc{linux} & Multi-stage attacks from reconnaissance to escalation& \textsc{wazuh} alerts\\
    CSLE-IDS-2024 & \textsc{linux} & Software exploits, e.g., \textsc{cve}-2015-1427&\textsc{snort} alerts \\
    \bottomrule
    \end{tabular}
    \caption{The four evaluation datasets, from \cite{GARCIA2014100}, \cite{ghorbani18dataset}, \cite{Wurzenberger24}, and \cite{kmi24feedback-control}, respectively, include diverse attacks, logs, and systems.}
    \label{tab:dataset}
\end{table*}

% \subsection{Online Lookahead Planning \& Action Generation}
% alert classification, meet threshold (GPT)
% fine-tune action generation (display loss)

% \label{subsec:planning_action_generation}

\paragraph{Calibration}
When a set of candidate tactics $\Theta$ is provided externally (e.g., from a frontier model such as \textsc{gpt}-5.2), we first inspect these tactics to ensure they are consistent with the logs.
For each tactic $\hat{\theta} \in \Theta$, we prompt the \textsc{llm} to produce a set of unique
\([\texttt{Classification}]\)--\([\texttt{Priority}]\) pairs and compare them against the labeled log text.
We compute a unique-pair precision (and recall) score and require the precision to exceed a threshold
$\tau_{\mathrm{AP}}$ (e.g., $\tau_{\mathrm{AP}}=0.6$) to proceed with multi-step planning. If no tactic's generation passes the threshold, we trigger the \emph{calibration} procedure rather than running a full multi-step lookahead. Table~\ref{tab:lora} presents the configuration of the action generation batch $N$ and trajectory batch $M$.

\subsection{Evaluation and Baselines}

Our experiment aggregates four evaluation datasets summarized in Table~\ref{tab:dataset}, which include a variety of attacks, alert logs, and system configurations. The recovery actions in these evaluation datasets are used to assess the effectiveness of the actions we generate. We rely on \textsc{gpt}-5.2 for such assessment, and measure the recovery time in discrete time units, assigning a time cost of 1 to all actions, and for those superfluous, less effective steps assessed by \textsc{gpt}-5.2, an additional cost of 1 is assigned as a kind of penalty. For actions generated that do not lead to a recovery terminal state, the time cost is set to 20.  

We compare our \textsc{llm} agent with three frontier models: \textsc{deepseek-r1} \cite{deepseek-r1}, \textsc{gemini 2.5 pro} \cite{gemini}, and \textsc{openai o3} \cite{openai}. We remind the reader that our model, with 14b parameters, is significantly lightweight and of the same size as the baseline from \cite{hammar2025incident}. 

%\tikzexternaldisable
\begin{figure}[h]
  \centering
%\tikzsetnextfilename{eval_bars_1}        
  \scalebox{0.81}{
   \begin{tikzpicture}
\node[scale=1] (kth_cr) at (0,3)
{
\begin{tikzpicture}
\begin{axis}[
   ybar,
    title style={align=center},
    ticks=both,
    ymin=0,
    axis x line = bottom,
    axis y line = left,
    axis line style={-|},
    enlarge y limits={lower, value=0.1},
    enlarge y limits={upper, value=0.22},
    xtick=\empty,
    ymajorgrids,
    xticklabels={},
    legend style={nodes={scale=0.8, transform shape}, at={(0.04, -0.4)}, align=left, anchor=west, legend columns=3, draw=none},
%    every axis legend/.append style={nodes={right}, inner sep = 0.2cm, scale=0.2, transform shape},
   x tick label style={align=center, yshift=-0.1cm},
    enlarge x limits=4,
    width=11cm,
    bar width=0.7cm,
    height=4cm,
    ]

\addplot+[
  draw=black, color=black,fill=Orange!90,
  nodes near coords,
  every node near coord/.append style={
      anchor=south,
      shift={(axis direction cs:0,3)}, 
      font=\small\bfseries, fill=white,
scale=0.78,
  },
  error bars/.cd, y dir=both, y explicit,
] coordinates {
  (0,10.30) +- (0,1.6733)
};
    
\addplot+[
  draw=black, color=black,fill=Red!90,
postaction={pattern=crosshatch},
  nodes near coords,
  every node near coord/.append style={
      anchor=south,
      shift={(axis direction cs:0,1.2)}, 
      font=\small\bfseries, fill=white,
scale=0.78,
  },
  error bars/.cd, y dir=both, y explicit,
] coordinates {
  (1,13.46) +- (0,1.09)
};

\addplot+[
  draw=black, color=black,fill=OliveGreen!60,
postaction={pattern=dots},
  nodes near coords,
  every node near coord/.append style={
      anchor=south,
      shift={(axis direction cs:0,1.6)}, 
      font=\small\bfseries, fill=white,
scale=0.85,
  },
  error bars/.cd, y dir=both, y explicit,
] coordinates {
  (2,16.21) +- (0,1.25)
};

\addplot+[
  draw=black, color=black,fill=bluetwo,
postaction={pattern=north east lines},
  nodes near coords,
  every node near coord/.append style={
      anchor=south,
      shift={(axis direction cs:0,1.9)}, 
      font=\small\bfseries, fill=white,
scale=0.82,
  },
  error bars/.cd, y dir=both, y explicit,
] coordinates {
  (3,17.28) +- (0,1.60)
};

\addplot+[
  draw=black, color=black,fill=Blue!40,
postaction={pattern=north west lines},
  nodes near coords,
  every node near coord/.append style={
      anchor=south,
      shift={(axis direction cs:0,1.9)}, 
      font=\small\bfseries, fill=white,
scale=0.82,
  },
  error bars/.cd, y dir=both, y explicit,
] coordinates {
  (4,17.09) +- (0,1.43)
};

\addplot+[
  draw=black, color=black,fill=Orange!90,
  nodes near coords,
  every node near coord/.append style={
      anchor=south,
      shift={(axis direction cs:0,0.4)}, 
      font=\small\bfseries, fill=white,
scale=0.85,
  },
  error bars/.cd, y dir=both, y explicit,
] coordinates {
  (50,3.00) +- (0,0.41)
};

\addplot+[
  draw=black, color=black,fill=Red!90,
postaction={pattern=crosshatch},
  nodes near coords,
  every node near coord/.append style={
      anchor=south,
      shift={(axis direction cs:0,0.4)}, 
      font=\small\bfseries, fill=white,
scale=0.85,
  },
  error bars/.cd, y dir=both, y explicit,
] coordinates {
  (51,3.00) +- (0,0.41)
};

\addplot+[
  draw=black, color=black,fill=OliveGreen!60,
postaction={pattern=dots},
  nodes near coords,
  every node near coord/.append style={
      anchor=south,
      shift={(axis direction cs:0,0.55)}, 
      font=\small\bfseries, fill=white,
scale=0.85,
  },
  error bars/.cd, y dir=both, y explicit,
] coordinates {
  (52,3.30) +- (0,0.49)
};

\addplot+[
  draw=black, color=black,fill=bluethree,
postaction={pattern=north east lines},
  nodes near coords,
  every node near coord/.append style={
      anchor=south,
      shift={(axis direction cs:0,0.8)}, 
      font=\small\bfseries, fill=white,
scale=0.85,
  },
  error bars/.cd, y dir=both, y explicit,
] coordinates {
  (53,4.21) +- (0,0.68)
};

\addplot+[
  draw=black, color=black,fill=Blue!40,
postaction={pattern=north west lines},
  nodes near coords,
  every node near coord/.append style={
      anchor=south,
      shift={(axis direction cs:0,0.7)}, 
      font=\small\bfseries, fill=white,
scale=0.85,
  },
  error bars/.cd, y dir=both, y explicit,
] coordinates {
  (54,4.48) +- (0,0.59)
};

\legend{\textsc{our agent}$\quad$, \cite{hammar2025incident}$\quad$, \textsc{gemini 2.5}$\quad$, \textsc{openai o3}$\quad$, \textsc{deepseek-r1}}
\end{axis}
\end{tikzpicture}
};

%\legend{\textsc{our method}$\quad$, \textsc{gemini 2.5}$\quad$, \textsc{openai o3}$\quad$, \textsc{deepseek-r1}}
%\end{axis}
%\end{tikzpicture}
%};

%\node[inner sep=0pt,align=center, scale=0.8, rotate=0, opacity=1] (obs) at (0.5,5)
%{
%Average
%};

\node[inner sep=0pt,align=center, scale=0.9, rotate=0, opacity=1] (obs) at (-2,2.25)
{
  Recovery time
};

\node[inner sep=0pt,align=center, scale=0.9, rotate=0, opacity=1] (obs) at (3,2.25)
{
  \% Failed recoveries
};

\end{tikzpicture}            
  }
  \caption{Evaluation results ($\downarrow$ better): comparison between our method and frontier \textsc{llm}s. Bar colors relate to different methods; bar groups indicate performance metrics; numbers and error bars indicate the mean and the standard deviation from $5$ evaluations with different random seeds.}
  \label{fig:eval_bars_1}
\end{figure}
% \tikzexternalenable

%\tikzexternaldisable
\begin{figure}[h]
  \centering
%\tikzsetnextfilename{eval_bars_2}
  \scalebox{0.81}{
   \begin{tikzpicture}

\node[scale=1] (kth_cr) at (0,-6.6)
{
\begin{tikzpicture}
\begin{axis}[
   ybar,
    title style={align=center},
    ticks=both,
    ymin=0,
    axis x line = bottom,
    axis y line = left,
    axis line style={-|},
    enlarge y limits={lower, value=0.1},
    enlarge y limits={upper, value=0.22},
    xtick=\empty,
    ymajorgrids,
    xticklabels={},
    legend style={nodes={scale=0.8, transform shape}, at={(0.24, 1.1)}, align=left, anchor=west, legend columns=4, draw=none},
%    every axis legend/.append style={nodes={right}, inner sep = 0.2cm, scale=0.2, transform shape},
   x tick label style={align=center, yshift=-0.1cm},
    enlarge x limits=1.5,
    width=11.5cm,
    bar width=0.9cm,
    height=4cm,
    ]

\addplot+[
  draw=black, color=black,fill=OliveGreen!70,
  nodes near coords,
  every node near coord/.append style={
      anchor=south,
      shift={(axis direction cs:0,1.8)}, 
      font=\small\bfseries, fill=white,
scale=0.8,
  },
  error bars/.cd, y dir=both, y explicit,
] coordinates {
  (0,10.30) +- (0,1.62)
};

\addplot+[
  draw=black, color=black,fill=OliveGreen!30,
postaction={pattern=dots},
  nodes near coords,
  every node near coord/.append style={
      anchor=south,
      shift={(axis direction cs:0,2)}, 
      font=\small\bfseries, fill=white,
scale=0.85,
  },
  error bars/.cd, y dir=both, y explicit,
] coordinates {
  (1,20.87) +- (0,1.85)
};

\addplot+[
  draw=black, color=black,fill=Blue!50,
  nodes near coords,
  every node near coord/.append style={
      anchor=south,
      shift={(axis direction cs:0,1.8)}, 
      font=\small\bfseries, fill=white,
scale=0.8,
  },
  error bars/.cd, y dir=both, y explicit,
] coordinates {
  (20,10.30) +- (0,1.62)
};
\addplot+[
  draw=black, color=black,fill=Blue!20,
postaction={pattern=dots},
  nodes near coords,
  every node near coord/.append style={
      anchor=south,
      shift={(axis direction cs:0,2.4)}, 
      font=\small\bfseries, fill=white,
scale=0.85,
  },
  error bars/.cd, y dir=both, y explicit,
] coordinates {
  (21,16.20) +- (0,1.80)
};

\addplot+[
  draw=black, color=black,fill=Red!70,
  nodes near coords,
  every node near coord/.append style={
      anchor=south,
      shift={(axis direction cs:0,1.8)}, 
      font=\small\bfseries, fill=white,
scale=0.8,
  },
  error bars/.cd, y dir=both, y explicit,
] coordinates {
  (40,10.30) +- (0,1.62)
};
\addplot+[
  draw=black, color=black,fill=Red!30,
postaction={pattern=dots},
  nodes near coords,
  every node near coord/.append style={
      anchor=south,
      shift={(axis direction cs:0,1.4)}, 
      font=\small\bfseries, fill=white,
scale=0.85,
  },
  error bars/.cd, y dir=both, y explicit,
] coordinates {
  (41,12.62) +- (0,2.33)
};

%\legend{with$\quad$, without}
\end{axis}
\node[inner sep=0pt,align=center, scale=0.8, rotate=0, opacity=1] (obs) at (0.8,2.6)
{
  Recovery time
};
\end{tikzpicture}
};

\node[inner sep=0pt,align=center, scale=0.75, rotate=0, opacity=1] (obs) at (-2.8,-8.1)
{
  Fine-tuning
};

\node[inner sep=0pt,align=center, scale=0.75, rotate=0, opacity=1] (obs) at (0.3,-8.1)
{
  Planning
};

\node[inner sep=0pt,align=center, scale=0.75, rotate=0, opacity=1] (obs) at (3.5,-8.1)
{
  Context adaptation
};

\end{tikzpicture}            
  }
\caption{Ablation-study results for the recovery time metric ($\downarrow$ better). Bar groups correspond to a specific step of our method; filled bars show performance with each step, and dotted bars show performance with the step removed; numbers and error bars indicate the mean and standard deviation across $5$ evaluations with different random seeds.}
  \label{fig:eval_bars_2}
\end{figure}
%\tikzexternalenable 
Figure~\ref{fig:eval_bars_1}  summarizes the major evaluation results. While they all share similar failure rates, our model leads to the shortest recovery time than others, as shown in Fig.~\ref{fig:eval_bars_1}, whereas those frontier models, even though equipped with the latest security knowledge base, do not create an effective response plan, as they are not tailored to the security task. 

To further evaluate the importance of each functionality, we conduct ablation studies removing the fine-tuning (i.e., Perception \& Reasoning functions), planning, and the in-context adaptation mechanism. Fig.~\ref{fig:eval_bars_2} compares the response performance before and after removing these key functionalities, which demonstrates that fine-tuning and planning play a significant role in the entire workflow, whereas the context adaptation, while still improving the performance, is not as instrumental as the other two. We speculate that the modest improvement is due to the fact that the test data points are mostly short sequences of recovery actions, typically 5 actions. Another extension is to evaluate our model on longer response processes, where the challenging longer-context situation can highlight the impact of our in-context adaptation scheme.   

We finally remark on the scalability of the proposed \textsc{llm} agent, which is the main limitation of our approach. The major computational expense is due to the Monte Carlo tree search and scales in $O(MN)$ time. All implementations are operated on Google Cloud with one A100 \textsc{gpu}. We observe that processing one incident takes, on average, 20 minutes to generate a five-action response plan. When deploying our agent on more complex network systems against more sophisticated tactics, the agent needs larger search trees, which soon render generation time disappointing.  One of the most pressing extensions is to develop cost-efficient simulation or parallel computing methods for prompt incident response. 

\section{Conclusion}
We present an in-context adaptive response planning method based on an \textsc{llm} agent, resulting in an end-to-end incident response workflow without the structured modeling and simplification used in prior work on \textsc{rl}-based incident response. We model response planning as a partially observed Markov decision process and integrate Perception, Reasoning, Planning, and Action functionalities into a single lightweight \textsc{llm} model inspired by the Monte-Carlo tree search method. We evaluate our \textsc{llm} agent on diverse incidence datasets against frontier \textsc{llm} models, and our \textsc{llm} agent achieves $23\%$ shorter recovery time than baselines. 

The most pressing extension is to address the scalability issue in our \textsc{llm}-based simulation, which can be time-consuming when deploying the agent to complex network environments. Another extension is to improve the current evaluation procedures by introducing more realistic time costs, comprehensive response action assessment, and long action sequences in the log data.

\bibliography{aaai25}
\begin{appendices}
\setcounter{secnumdepth}{2} % Re-enables numbering for sections and subsections
\renewcommand{\thesection}{\Alph{section}} % Sets numbering 

\end{appendices}

\end{document}